\documentclass[mathleft
]{an}
\usepackage{graphicx}
\usepackage[authoryear]{natbib}
\usepackage{times}
\def\simlt{\mathrel{\hbox{\rlap{\hbox{\lower4pt\hbox{$\sim$}}}\hbox{$<$}}}}
\def\simgt{\mathrel{\hbox{\rlap{\hbox{\lower4pt\hbox{$\sim$}}}\hbox{$>$}}}}

\newcommand       \aap          {A\&A }
\newcommand       \ssr           {Spac.Scienc.Review}
\newcommand       \aapr         {Astronomy and Astrophysics Review}
\newcommand       \nat         {Nature}

\bibpunct{(}{)}{;}{a}{}{,}
\begin{document}

% The following seven commands are intended for editorial usage and should be ignored by
% the author(s).
\Pagespan{1}{6}% Document's page range. 
% If second parameter is left empty, the last page is computed automatically.
\Yearpublication{2013}%
\Yearsubmission{2013}%
\Month{07}%   
\Volume{xxx}%  
 \Issue{xx}% 
% \DOI{This.is/not.aDOI}% 

\title{Review of the Theory of Pulsar Wind Nebulae}

\author{N.~Bucciantini\inst{1,2}\fnmsep\thanks{Corresponding author:
  \email{niccolo@arcetri.astro.it}\newline}
}
\titlerunning{Review of the Theory of PWNe}
\authorrunning{N.~Bucciantini}
\institute{INAF, Osservatorio Astrofisico di Arcetri, L.go E. Fermi 5,
  50125, Firenze, Italy
\and 
INFN, Sezione di Firenze, via G. Sansone 1, 50019, Sesto Fiorentino (FI), Italy
}

\received{30 July 2013}
\accepted{25 September 2013}
\publonline{later}

\keywords{pulsars: general - magnetohydrodynamics -
  radiation mechanism: non-thermal  - stars: winds, outflows -  supernova remnants }

\abstract{
  Pulsar Wind Nebulae (PWNe) are ideal astrophysical laboratories where high energy relativistic
phenomena can be investigated. They are close, well resolved in our
observations, and the knowledge
 derived in their study has a strong impact in many other fields, from AGNs to GRBs. Yet
there are still unresolved issues, that prevent us from a full clear understanding of these objects.
The lucky combination of high resolution X-ray imaging and numerical codes to handle the outflow
and dynamical properties of relativistic MHD, has opened a new avenue of investigation that has
lead to interesting progressed in the last years. Despite all of these, we do not understand yet
how particles are accelerated, and the functioning of the pulsar wind and pulsar magnetosphere,
that power PWNe. I will review what is now commonly known as the MHD
paradigm, and in particular I will focus on various approaches that
have been and are currently used to model these systems. For each I
will highlight its advantages and limitations, and degree of applicability.
}

\maketitle

\section{Introduction}

Pulsar Wind Nebulae (PWNe) form when the ultra-relativistic wind
from a pulsar interacts with the ambient medium, either the parent
Supernova Remnant (SNR) or the
ISM \citep{Chevalier04a,Gaensler_Slane06a,Bucciantini08a}.
They are bubbles of relativistic particles, pairs and perhaps ions,
emitting in a broad band spectrum from radio to $\gamma$-rays, via
synchrotron and Inverse Compton emission \citep{Weiler_Panagia78a,Asaoka_Koyama90a,Harding96a,Bandiera00a,Kargaltsev_Pavlov10a}. The best example of a PWN, 
often considered the prototype of this entire class of objects, is the
Crab Nebula \citep{Hester08a}. It is such a well known system, that the majority of our
models are tuned to address its properties.   

There are two key ingredients that go into any good recipe for
cooking a realistic model of a PWN. The first ingredient is the
acceleration mechanism responsible for producing the non-thermal
particle's distribution function that is observed. The second
ingredient is the dynamical model of the PWN, describing the advection
and evolution of the flow within these systems.

The problem of accelerating the observed particle's distribution function is a
long standing one \citep{Arons_Tavani94a,Bykov_treumann11a}. Recent
{\it particle-in-cell} simulations, in a variety of different
conditions, have not helped to figure out a satisfactory
solution. Ions have been invoked, but they must carry a substantial
fraction of the total spin-down energy, for which there is no clear
evidence \citep{Amato_Arons06a,Sironi_Spitkovsky11b}. Efficient {\it Fermi-like} acceleration of a pure pair plasma
requires very small magnetizations in the upstream pulsar wind, that
are reached only in a small solid angle \citep{Sironi_Spitkovsky09a}. Reconnection of a striped
wind \citep{Lyubarsky03a,Lyubarsky_Livers08a,Sironi_Spitkovsky11a}, can provide hard spectra, compatible with the radio emission, but
requires very high multiplicity and small Lorentz factors, contrasting
the expectations for pulsar wind dynamics \citep{Goldreich_Julian69a,Bogovalov97a,Hibschman_Arons01a}. There is also the
possibility of continuous distributed turbulent acceleration in the bulk of the
nebula \citep{Barnes_Scargle73a,Bucciantini_Arons+11a}. There is however a general consensus, based on the wavelength
dependent size of these objects, that X-ray emitting particles must be
accelerated at the wind termination shock. The origin of radio
emitting ones is still currently debated
(\citealt{Atonyan_Aharonian96a}; Olmi et al. Private communication).

On the other hand dynamical models for the flow properties have
developed to a greater success. There is now a well established
paradigm, based on MHD. The key assumption is that the Larmor radius
of the vast majority of the particles is much smaller that the
typical size of these  nebulae. This is true up to the highest
observed energies. This allows one to use a fluid description, that can easily be
formulated in terms of MHD. Once coupled with a prescription for
particle's acceleration, it can easily be turned into an emission
model. In principle, a satisfactory model, built along these lines,
should be able to relate the observed morphology to the pulsar wind
properties, and to provide a unifying picture for all these objects.

Recently, the tools and approches that have been used in the
investigation of PWNe have been applied to Long and Short
GRBs \citep{Dall'Osso-Stratta+11a,Bucciantini-Quataert+08a,Bucciantini-Quataert+09a,Bucciantini-Metzger+12a}. Indeed these systems too can be modeled as expanding
relativistic magnetized bubbles, and share a large amount of physics
and dynamics with PWNe. 

I will review in the following the various approaches that have been
presented in the literature to model these systems. I will start from
the simpler 1-Zone and 1-D models and show how the extension to higher
dimensions can both provide a more accurate description and  also
raise new questions. I will conclude with a brief review of various
models for pulsar bow-shock nebulae.

\section{1-Zone Models}

With the label {\it 1-Zone  model} (or equivalently  {\it 0+1 model}), we refer to any model where the
PWN is described as an object, whose geometry is only characterized by a typical size
(often its radius), which in principle  can also depend on time. These
models completely neglect the possibility of internal structure in the
PWN and assume that all quantities are uniform. Despite the
simplicity of this approach, these models have proved very successful
for the investigation of the global rough properties of these
systems. In particular they have been applied to the study of the
evolution of PWNe, often in conjunction with their interaction with the
partent SNR. The first 1-Zone model for PWNe was put forward by
\citet{Pacini_Salvati73a}.  Here the evolution of a non-thermal distribution of
particles, inside an expanding bubble, subject to adiabatic losses,
radiation losses and a continuous injection, was investigated, in
relation to the non-thermal emission of the Crab Nebula. All the key ideas
of 1-Zone models were already laid down. The analytical approach of
this seminal work however,  required assumptions on the temporal
evolution of the various quantities in the PWN that are reasonable only
for young systems in the so called {\it free-expansion phase} (the first few
thousands years). It moreover imposes constraints on the kind of
injected particle's distribution (a power-law). This model has been
further extended to take partially into account the interaction with the
SNR \citep{Reynolds_Chevalier84a,Reynolds_Chanan84a}, or changes in
injection history \citep{Atoyan99a}. More recently numerical models
have been developed
\citep{Bednarek_Bartosik03a,Torres_Cillis+13a,Tanaka_Takahara13a,Tanaka_Takahara10a,Matin_Torres+12a,Gelfand_Slane+09a,Bucciantini_Arons+11a}. Using
a numerical approach, the evolution can be followed for a longer time taking into
account the various phases of the interaction with the parent SNR. 
These models allow us, not only to include a complex evolutionary
history, but also to model the evolution of distribution functions,
that are not necessarily a single power law (either broken power laws
have been considered, or a power-law plus a Maxwellian).  In
particular, the work by \citet{Gelfand_Slane+09a} is exemplary for showing the richness of behaviors that is found even within the 1-Zone treatment, and the degree of complexity that one can investigate using simple tools.

The main advantage of 1-Zone models is their easiness of
implementation, and the fact that they allow us to sample a large
parameter space of often unknown injection or evolutionary
properties. Of course one should not expect more than a rough
quantitative agreement with the global observed properties of these
system. In particular these models are not reliable for the high-energy
X-ray emission properties, because, as it will be shown later, they are
strongly affected by the internal flow dynamics, that these kinds of
models fail completely to consider. On the other hand the Radio
emission properties, the location of spectral breaks, the ratio of
synchrotron to Inverse Compton emission, can be fairly well
modelled. The same holds for the overall evolution, where the typical
duration and timescale of the various phases of interaction with the
SNR are correctly recovered.

Given that quite often the quality of the available data is limited,
1-Zone models turn out to be perhaps the only justifiable approach. On the other hand they
also constitute a good starting point for more accurate,
multidimensional studies.

\section{1-D Models}

It is a general expectation that, increasing the dimensionality of a
model, should lead to a more realistic description of a system: a
description closer to the true working. If one includes also a simple spatial dependence of the various
quantities (often a dependence on the radial coordinate from the
pulsar) it is possible to develop {\it 1-D models}  (or equivalently
  {\it 1+1 models}).  The first 1-D model of the Crab
Nebula was put forward by \citet{Rees_Gunn73a}. Already this simple description
allows one to understand some of the spatial features/properties of
this object. The under-luminous region, centered on the location of
the pulsar, is interpreted as the ultra-relativistic unshocked
wind. The wavelength dependent size of the Crab Nebula, with radio
emission extending further outside the X-ray one, is explained in term
of combined bulk flow from the termination shock to the outer edge of
the nebula, and synchrotron cooling. The existence of a bright torus
is understood as due to compression of the magnetic field in the outer
layers of the nebula. This model also showed that it was in principle
possible to use the nebular properties to infer the conditions of the
pulsar wind, like its Lorentz factor and/or magnetization. This model has
been further developed by \citet{Kennel_Coroniti84a,Kennel_Coroniti84b} who solved the full set of
relativistic MHD equations, and provided a quantitative estimate of
the structural properties of the Crab Nebula. Later \citet{Emmering_Chevalier87a}
presented a time dependent analytical solution. 

Further development of 1-D models came mainly from numerical
investigation, which allowed us to avoid many of the simplifications
required by analytical treatment. \citet{Blondin-Chevalier+01a} and \citet{van_der_Swaluw-Achterberg+01a} were the first to investigate the dynamics
of the PWN-SNR interaction using a hydrodynamical code. All the various
phases of the mutual interaction where followed: the {\it
  free-expansion phase} \citep{Chevalier-Fransson92a}, the {\it reverberation phase}, and
the final {\it Sedov phase}. Similar works have also been carried out 
by \citet{Bucciantini-Blondin+03a,Bucciantini-Bandiera+04a} which were the first to use relativistic
MHD and to investigate the role of magnetic field. More recently
\citet{de_Jager-Fereira+08a} have used a 1-D model to compute the evolution of an
injected distribution function, and to compute an integrated spectrum
for G21.5-0.9. Latest results have been presented by \citet{Vorster-Ferreira+13a}.

One of the major issues raised by 1-D models is the so called {\it $\sigma$
  problem} \citep{Melatos98a}. Traditionally the  {\it $\sigma$
  problem} mixes together two different problems: one related to the
efficiency of accelerating particles at a shock, the other to the
ability of slowing down a relativistic magnetized outflow. The latter is a
dynamical problem, and in the following I will only consider this
aspect. 
In Ideal MHD a radial highly-magnetized (with a toroidal field) supersonic
relativistic pulsar wind cannot be slowed down by a shock to match
the typical expansion speeds of the confining SNR. For this to happen
it is required a small value of the magnetization \citep{Arons04a}. In fact, if the ratio of Poynting
flux to total energy flux in the pulsar wind exceeds $\sim 10^{-3}$, a
quasi-stable shock cannot form, and one expects a subsonic solution
for the pulsar outflow, a {\it breeze}. This contrasts strongly the
expectations of pulsar wind electrodynamics \citep{Michel69a,Goldreich-Julian70a,Beskin-Kuznetsova+98a,Bogovalov01a}, which suggest that the
magnetization in the wind, at best, with minor deviations from radial outflow, could be $\sim 1$ at the typical
distance of the nebulae \citep{Chiueh_Li+98a}. A closer issue is the inability to reproduce
the relative size of X-ray an Radio in the Crab Nebula, and similar
young systems \citep{Amato-Salvati+00a}, and the correct softening of the spectrum
with the distance from the pulsar \citep{Reynolds04a}. All of these problems
are related to the fact that in 1-D models there is a one-to-one
correspondence between distance, age, magnetic field and flow
velocity. This inevitably leads to onion-like structures, which are
the exact opposite of the complete mixing at the base of 1-Zone models.

\section{2-D Models}

It is evident that, in the presence of a toroidal magnetic field,
stresses will arise in the nebula, leading to a prolate
shape. \citet{Begelman-Li92a} were the first to develop a 2-D model accounting
for this effect, in an attempt to explain the observed prolate
geometry of the Crab Nebula, and to use it as a further constrain on
the plasma magnetization. Models that assume a cylindrical geometry
are referred as {\it 2-D models}  (or equivalently
  {\it 2+1 models}).

However the interest in multidimensional models for PWNe received a
considerable boost only at the start of the 21st century. Optical and X-ray
images from HST, CHANDRA and XMM-Newton have shown that the inner
region of PWNe is characterized by a complex axisymmetric structure,
generally referred as {\it jet-torus structure}. First observed in the
Crab \citep{Hester-Scowen+95a,Weisskopf-Hester+00a}, it has been
detected in many other PWNe \citep{Gaensler-Arons+02a,Lu-Wang+02a,Camilo-Gaensler+04a,Pavlov-Teter+03a,Romani-Ng03a,Romani-Ng+05a,Slane-Helfand+04a}, and there is
a general consensus that such features are always present in young
systems. However given the complex dynamics, associated with the
larger degrees of freedom due to the increased dimensionality, only the
use of numerical codes for computational fluid dynamics
\citep{Komissarov99a,Del_Zanna-Bucciantini+03a}
 allowed us to
move beyond the simple analytical solution by
\citet{Begelman-Li92a}. \citet{van_der_Swaluw03a} was the first to present a numerical model
of a prolate nebula applied to 3C58. It was however the investigation
of the dynamics of the pulsar wind that drove most of the successive
attempts at modeling the inner jet-torus structure. Theoretical \citep{Begelma-Li94a,Beskin-Kuznetsova+98a}
and numerical
\citep{Contopoulos-Kazanas+99a,Bogovalov01a,Timokhin06a,Komissarov06a,Spitkovsky06a,Bucciantini-Thompson+06a}
studies of relativistic winds from
pulsars agree that, at the typical distance of the termination shock,
the wind luminosity and magnetization are not uniform, but can be
described by the so called {\it split monopole} solution: most of the
energy is confined to the equatorial region, and there is no evidence
of jets or collimated pulsar outflows along the axis. It is
reasonable to expect that such a wind could drive a complex dynamics at least in
the inner region. \citet{Bogovalov-Khangoulian02a,Bogovalov-Khangoulian02b} were the first to investigate this
problem, modeling the flow dynamics at the termination shock. Later
\citet{Lyubarsky02a} suggested that hoop stresses in the body of the nebula
could lead to the collimation of the jet. This has been extensively confirmed
numerically by different groups \citep{Komissarov-Lyubarsky04a,Del_Zanna-Amato+04a,Bogovalov-Chechetkin+05a,Del_Zanna-Volpi+06a,Volpi-Del_Zanna+08a}. This numerical work has
proved quite rich in its results, and various aspects of the nebular
morphology have been investigated, from the details of emission maps
\citep{Del_Zanna-Volpi+06a}, to polarization \citep{Bucciantini-Del_Zanna+05a}, to the overall spectrum
\citep{Volpi-Del_Zanna+08a}. We are now in a position where 2-D models can accurately
reproduce the observed X-ray features of the Crab Nebula, including
the wisps and the knot, in terms of relative size and brightness. More
recently attention has focused of the problem of time
variability. Variability in the X-ray was known since the first
observations. In particular the variability in the wisps region \citep{Hester-Scowen+95a,Schweizer-Bucciantini+13a}. 2-D
models have shown that such variability can be reproduced \citep{Bucciantini-Del_Zanna06a,Bucciantini08a,Camus-Komissarov+09a}
both in term of its typical timescale and in its morphological
pattern, of outgoing waves. Interestingly these 2-D models have shown
that the limiting value of the wind magnetization $\sigma$, required to
explain the observed morphology and the presence of a jet, is about
one order of magnitude higher than the maximum value allowed in 1-D models, and this mostly
because the extra degree of freedom of the 2-D geometry allows to accommodate more
magnetized winds. There is also a large degree of internal fluid
turbulence and mixing. This strongly contrasts the basic assumption of 1-D models
(the onion-like structure), and is suggestive that reality could be
closer  to the full mixing
assumption of 1-Zone models.

Apart from the dynamics of the internal region, 2-D models have been
also applied to the interaction with the SNR. \citet{Jun98a} and  \citet{Bucciantini-Amato+04a} were the
first to investigate the Rayleigh-Taylor instability at the interface
with the SNR ejecta, both in the magnetized an unmagnetized case,
showing the development of the filamentary network observed in the
Crab Nebula. \citet{Blondin-Chevalier+01a} were the first to investigate the 2-D dynamics
of the reverberation phase, showing the high level of mixing between
the SNR and PWN, and the role of density gradients in the ISM, leading
to a relic PWN displaced with respect to the pulsar, as it is often
observed in old systems. \citet{van_der_Swaluw-Downes+04a} investigated in detail the
interaction of a pulsar moving across the SNR, the transition from
spherical PWNe in the early phase to a cometary structure, and in
\citet{van_der_Swaluw-Achterberg+02a} the interaction with the SNR forward shock was also taken
into account. \citet{Ferreira-De_Jager08a} analyzed the role of ISM density gradients,
in the presence of an ordered magnetic field.  

\section{3-D Models}

It has been known for a long time
\citep{Begelman98a,Nalewajko-Begelman12a} that configurations with a
purely toroidal magnetic  field are subject to current driven instability ({\it
  kink instability}). Indeed a kink-like instability is seen in PWNe
\citep{Pavlov-Teter+03a,Mori-Burrows+04a}.
To fully capture the dynamics associated with
this instability it is necessary to develop full {\it 3-D models} (also
known as {\it 3+1 models}). Recently this process has been
studied via numerical simulations in the simplified regime of a
magnetized column confined by a hot atmosphere \citep{O'Neill-Beckwith+12a,Mizuno-Lyubarsky+11a,Mizuno-Lyubarsky+12a}. These simulations have
shown that the instability reaches the non-linear regime on a typical
Alfv\`enic timescale, and that, after about 10 Alfv\`enic timescales, the
original toroidal field has almost completely vanished. While this
suggests a very efficient way out of the $\sigma$ problem \citep{Begelman98a}, on the other
hand it is obvious that such a violent instability cannot operate in
PWNe, because polarization measures \citep{Smith-Jones+88a,Velusamy85a,Bietenholz91a} constrain the level of ordered
magnetic field to be very high. Part of the reason for this
inconsistency can be found in the simplified regime that these models
consider. In PWNe the toroidal magnetic field is continuously injected by
the pulsar wind, and the PWN is not pressure confined but confined by a
high density wall at the contact discontinuity with the SNR ejecta. On
one hand this implies that there must be a balance between the
injection and dissipation of the toroidal field, on the other hand, it implies
that in the inner region, close to the termination shock, where the
plasma is injected, the magnetic field
is bound to be close to toroidal, while instabilities would only act in
the outer regions. Unfortunately a full 3-D simulation of a PWN,
evolving long
enough to reach a balance, has not been carried out yet, mostly
because of computational cost.  

However recently \cite{Porth-Komissarov+13a} presented a very interesting preliminary
study, where the dynamics of the PWN in a full 3-D setting was
investigated for a time-length corresponding to 70 yr. While this is
about one order of magnitude smaller than the age of the Crab Nebula, and it
can be debated if the nebula has reached a state of dynamical
equilibrium, nevertheless this timescale is longer that a typical
Alfv\`enic crossing time, and longer than the synchrotron cooling time for
X-ray emitting particles. The results confirm the qualitative expectations: the
inner region close to the termination shock is still dominated by a toroidal field, and
preserves the axisymmetric structure of 2D simulations; a jet is formed
and extends in the body of the nebula; instabilities at the outer edge of
the nebula act to reduce the net toroidal field component. It is also
found that the nebula can accommodate even a wind with high $\sigma$, of order
unity. Recalling that, even minor deviations of the wind from a purely radial
flow, can lower $\sigma$ to values of order unity, at the typical
distance of the termination shock radius \citep{Chiueh_Li+98a}, this suggests that the $\sigma$ problem is likely an artifact of the
1-D geometry of theories of old. However this does not rule out the
possibility of other non-ideal effects, like dissipation and/or
reconnection to play a role.

\section{Bow-Shocks}

As we pointed out in the previous sections, few works have been devoted to
the multi-dimensional investigation of the dynamical evolution of old
PWNe inside the SNR. Interestingly there are several works focusing on
the interaction with the ISM. Pulsars moving in the ISM produce the so
called {\it pulsar wind bow-shock nebulae}
\citep{Cordes-Romani+93a,Bell-Bailes+95a,Gaensler-van_der_Swaluw+04a,Pavlov-Sanwal+06a,Kargaltsev-Misanovic+08a,Ng-Bucciantini+12a}. The
first attempt to model the physics
 of these
systems, accounting for the presence of neutral Hydrogen in the ISM, was done by \citet{Bucciantini-Bandiera01a,Bucciantini02b} extending the {\it thin-layer
  approximation} \citep{Giuliani82a} used to model cometary nebulae
\citep{Bandiera93a,Wilkin96a,Wilkin00a}. The {\it thin layer} approximation is
conceptually analogous
to a 1-D model, given that it neglects the thickness of the nebula,
and all quantities depend only on the distance from the apex. These
models however provide a quite good description of the head region of
these nebulae in terms of shape, thickness to hydrogen penetration,
H$_\alpha$ luminosity, as was later confirmed by more accurate 2-D
axisymmetric simulations both in the hydrodynamical regime \citep{Bucciantini02a,Gaensler-van_der_Swaluw+04a}
and in the relativistic MHDl regime
\citep{Bucciantini-Amato+05a}. A 3-D extension of the study of these
systems to take into account either a non-uniform ambient medium, or the
anisotropy in the pulsar wind energy flux (in analogy with 2-D
simulations of Crab-like PWNe) has been presented by \citet{Vigelius-Melatos+07a}.

More recently bow-shock models have also been developed for the
interaction and confinement of the pulsar wind
in binary systems
\citep{Bogovalov-Khangulyan+08a,Bogovalov-Khangulyan+12a,Bosch-Ramon-Barkov+12a}. In
this case it is the wind from the companion that provides the confining
medium. These models have been used to explain orbital variations and modulations of the
high energy comptonized emission.

\acknowledgements
The author wishes to thank his many collaborators, among whom: Luca Del Zanna,
Elena Amato, Jonathan Arons, Rino Bandiera, Serguei Komissarov, that
have all substantially contributed to the development of this field.

%\bibliography{my}{}

\begin{thebibliography}{118}
\providecommand{\natexlab}[1]{#1}

\bibitem[{{Amato} and {Arons}(2006)}]{Amato_Arons06a}
{Amato}, E. and {Arons}, J.: 2006, \apj 653, 325.

\bibitem[{{Amato} et~al.(2000){Amato}, {Salvati}, {Bandiera}, {Pacini}, and
  {Woltjer}}]{Amato-Salvati+00a}
{Amato}, E., {Salvati}, M., {Bandiera}, R., {Pacini}, F., and {Woltjer}, L.:
  2000, \aap 359, 1107.

\bibitem[{{Arons}(2004)}]{Arons04a}
{Arons}, J.: 2004, in F.~{Camilo} and B.M. {Gaensler}, eds., Young Neutron
  Stars and Their Environments, vol. 218 of \emph{IAU Symposium}, 163.

\bibitem[{{Arons} and {Tavani}(1994)}]{Arons_Tavani94a}
{Arons}, J. and {Tavani}, M.: 1994, \apjs 90, 797.

\bibitem[{{Asaoka} and {Koyama}(1990)}]{Asaoka_Koyama90a}
{Asaoka}, I. and {Koyama}, K.: 1990, \pasj 42, 625.

\bibitem[{{Atoyan}(1999)}]{Atoyan99a}
{Atoyan}, A.M.: 1999, \aap 346, L49.

\bibitem[{{Atoyan} and {Aharonian}(1996)}]{Atonyan_Aharonian96a}
{Atoyan}, A.M. and {Aharonian}, F.A.: 1996, \mnras 278, 525.

\bibitem[{{Bandiera}(1993)}]{Bandiera93a}
{Bandiera}, R.: 1993, \aap 276, 648.

\bibitem[{{Bandiera}(2000)}]{Bandiera00a}
{Bandiera}, R.: 2000, in P.C.H. {Martens}, S.~{Tsuruta}, and M.A. {Weber},
  eds., Highly Energetic Physical Processes and Mechanisms for Emission from
  Astrophysical Plasmas, vol. 195 of \emph{IAU Symposium}, 189.

\bibitem[{{Barnes} and {Scargle}(1973)}]{Barnes_Scargle73a}
{Barnes}, A. and {Scargle}, J.D.: 1973, \apj 184, 251.

\bibitem[{{Bednarek} and {Bartosik}(2003)}]{Bednarek_Bartosik03a}
{Bednarek}, W. and {Bartosik}, M.: 2003, \aap 405, 689.

\bibitem[{{Begelman}(1998)}]{Begelman98a}
{Begelman}, M.C.: 1998, \apj 493, 291.

\bibitem[{{Begelman} and {Li}(1992)}]{Begelman-Li92a}
{Begelman}, M.C. and {Li}, Z.Y.: 1992, \apj 397, 187.

\bibitem[{{Begelman} and {Li}(1994)}]{Begelma-Li94a}
{Begelman}, M.C. and {Li}, Z.Y.: 1994, \apj 426, 269.

\bibitem[{{Bell} et~al.(1995){Bell}, {Bailes}, {Manchester}, {Weisberg}, and
  {Lyne}}]{Bell-Bailes+95a}
{Bell}, J.F., {Bailes}, M., {Manchester}, R.N., {Weisberg}, J.M., and {Lyne},
  A.G.: 1995, \apjl 440, L81.

\bibitem[{{Beskin} et~al.(1998){Beskin}, {Kuznetsova}, and
  {Rafikov}}]{Beskin-Kuznetsova+98a}
{Beskin}, V.S., {Kuznetsova}, I.V., and {Rafikov}, R.R.: 1998, \mnras 299, 341.

\bibitem[{{Bietenholz}(1991)}]{Bietenholz91a}
{Bietenholz}, M.F.: 1991, {A Radio Study of the Crab Nebula.}, Ph.D. thesis,
  UNIVERSITY OF TORONTO (CANADA).

\bibitem[{{Blondin} et~al.(2001){Blondin}, {Chevalier}, and
  {Frierson}}]{Blondin-Chevalier+01a}
{Blondin}, J.M., {Chevalier}, R.A., and {Frierson}, D.M.: 2001, \apj 563, 806.

\bibitem[{{Bogovalov}(1997)}]{Bogovalov97a}
{Bogovalov}, S.V.: 1997, \aap 327, 662.

\bibitem[{{Bogovalov}(2001)}]{Bogovalov01a}
{Bogovalov}, S.V.: 2001, \aap 371, 1155.

\bibitem[{{Bogovalov} et~al.(2005){Bogovalov}, {Chechetkin}, {Koldoba}, and
  {Ustyugova}}]{Bogovalov-Chechetkin+05a}
{Bogovalov}, S.V., {Chechetkin}, V.M., {Koldoba}, A.V., and {Ustyugova}, G.V.:
  2005, \mnras 358, 705.

\bibitem[{{Bogovalov} and {Khangoulian}(2002a)}]{Bogovalov-Khangoulian02a}
{Bogovalov}, S.V. and {Khangoulian}, D.V.: 2002a, \mnras 336, L53.

\bibitem[{{Bogovalov} and {Khangoulyan}(2002b)}]{Bogovalov-Khangoulian02b}
{Bogovalov}, S.V. and {Khangoulyan}, D.V.: 2002b, Astronomy Letters 28, 373.

\bibitem[{{Bogovalov} et~al.(2012){Bogovalov}, {Khangulyan}, {Koldoba},
  {Ustyugova}, and {Aharonian}}]{Bogovalov-Khangulyan+12a}
{Bogovalov}, S.V., {Khangulyan}, D., {Koldoba}, A.V., {Ustyugova}, G.V., and
  {Aharonian}, F.A.: 2012, \mnras 419, 3426.

\bibitem[{{Bogovalov} et~al.(2008){Bogovalov}, {Khangulyan}, {Koldoba},
  {Ustyugova}, and {Aharonian}}]{Bogovalov-Khangulyan+08a}
{Bogovalov}, S.V., {Khangulyan}, D.V., {Koldoba}, A.V., {Ustyugova}, G.V., and
  {Aharonian}, F.A.: 2008, \mnras 387, 63.

\bibitem[{{Bosch-Ramon} et~al.(2012){Bosch-Ramon}, {Barkov}, {Khangulyan}, and
  {Perucho}}]{Bosch-Ramon-Barkov+12a}
{Bosch-Ramon}, V., {Barkov}, M.V., {Khangulyan}, D., and {Perucho}, M.: 2012,
  \aap 544, A59.

\bibitem[{{Bucciantini}(2002{\natexlab{a}})}]{Bucciantini02a}
{Bucciantini}, N.: 2002{\natexlab{a}}, \aap 387, 1066.

\bibitem[{{Bucciantini}(2002{\natexlab{b}})}]{Bucciantini02b}
{Bucciantini}, N.: 2002{\natexlab{b}}, \aap 393, 629.

\bibitem[{{Bucciantini}(2008)}]{Bucciantini08a}
{Bucciantini}, N.: 2008, in C.~{Bassa}, Z.~{Wang}, A.~{Cumming}, and V.M.
  {Kaspi}, eds., 40 Years of Pulsars: Millisecond Pulsars, Magnetars and More,
  vol. 983 of \emph{American Institute of Physics Conference Series}, 186--194.

\bibitem[{{Bucciantini} et~al.(2004{\natexlab{a}}){Bucciantini}, {Amato},
  {Bandiera}, {Blondin}, and {Del Zanna}}]{Bucciantini-Amato+04a}
{Bucciantini}, N., {Amato}, E., {Bandiera}, R., {Blondin}, J.M., and {Del
  Zanna}, L.: 2004{\natexlab{a}}, \aap 423, 253.

\bibitem[{{Bucciantini} et~al.(2005{\natexlab{a}}){Bucciantini}, {Amato}, and
  {Del Zanna}}]{Bucciantini-Amato+05a}
{Bucciantini}, N., {Amato}, E., and {Del Zanna}, L.: 2005{\natexlab{a}}, \aap
  434, 189.

\bibitem[{{Bucciantini} et~al.(2011){Bucciantini}, {Arons}, and
  {Amato}}]{Bucciantini_Arons+11a}
{Bucciantini}, N., {Arons}, J., and {Amato}, E.: 2011, \mnras 410, 381.

\bibitem[{{Bucciantini} and {Bandiera}(2001)}]{Bucciantini-Bandiera01a}
{Bucciantini}, N. and {Bandiera}, R.: 2001, \aap 375, 1032.

\bibitem[{{Bucciantini} et~al.(2004{\natexlab{b}}){Bucciantini}, {Bandiera},
  {Blondin}, {Amato}, and {Del Zanna}}]{Bucciantini-Bandiera+04a}
{Bucciantini}, N., {Bandiera}, R., {Blondin}, J.M., {Amato}, E., and {Del
  Zanna}, L.: 2004{\natexlab{b}}, \aap 422, 609.

\bibitem[{{Bucciantini} et~al.(2003){Bucciantini}, {Blondin}, {Del Zanna}, and
  {Amato}}]{Bucciantini-Blondin+03a}
{Bucciantini}, N., {Blondin}, J.M., {Del Zanna}, L., and {Amato}, E.: 2003,
  \aap 405, 617.

\bibitem[{{Bucciantini} and {Del Zanna}(2006)}]{Bucciantini-Del_Zanna06a}
{Bucciantini}, N. and {Del Zanna}, L.: 2006, \aap 454, 393.

\bibitem[{{Bucciantini} et~al.(2005{\natexlab{b}}){Bucciantini}, {del Zanna},
  {Amato}, and {Volpi}}]{Bucciantini-Del_Zanna+05a}
{Bucciantini}, N., {del Zanna}, L., {Amato}, E., and {Volpi}, D.:
  2005{\natexlab{b}}, \aap 443, 519.

\bibitem[{{Bucciantini} et~al.(2012){Bucciantini}, {Metzger}, {Thompson}, and
  {Quataert}}]{Bucciantini-Metzger+12a}
{Bucciantini}, N., {Metzger}, B.D., {Thompson}, T.A., and {Quataert}, E.: 2012,
  \mnras 419, 1537.

\bibitem[{{Bucciantini} et~al.(2008){Bucciantini}, {Quataert}, {Arons},
  {Metzger}, and {Thompson}}]{Bucciantini-Quataert+08a}
{Bucciantini}, N., {Quataert}, E., {Arons}, J., {Metzger}, B.D., and
  {Thompson}, T.A.: 2008, \mnras 383, L25.

\bibitem[{{Bucciantini} et~al.(2009){Bucciantini}, {Quataert}, {Metzger},
  {Thompson}, {Arons}, and {Del Zanna}}]{Bucciantini-Quataert+09a}
{Bucciantini}, N., {Quataert}, E., {Metzger}, B.D., {Thompson}, T.A., {Arons},
  J., and {Del Zanna}, L.: 2009, \mnras 396, 2038.

\bibitem[{{Bucciantini} et~al.(2006){Bucciantini}, {Thompson}, {Arons},
  {Quataert}, and {Del Zanna}}]{Bucciantini-Thompson+06a}
{Bucciantini}, N., {Thompson}, T.A., {Arons}, J., {Quataert}, E., and {Del
  Zanna}, L.: 2006, \mnras 368, 1717.

\bibitem[{{Bykov} and {Treumann}(2011)}]{Bykov_treumann11a}
{Bykov}, A.M. and {Treumann}, R.A.: 2011, \aapr 19, 42.

\bibitem[{{Camilo} et~al.(2004){Camilo}, {Gaensler}, {Gotthelf}, {Halpern}, and
  {Manchester}}]{Camilo-Gaensler+04a}
{Camilo}, F., {Gaensler}, B.M., {Gotthelf}, E.V., {Halpern}, J.P., and
  {Manchester}, R.N.: 2004, \apj 616, 1118.

\bibitem[{{Camus} et~al.(2009){Camus}, {Komissarov}, {Bucciantini}, and
  {Hughes}}]{Camus-Komissarov+09a}
{Camus}, N.F., {Komissarov}, S.S., {Bucciantini}, N., and {Hughes}, P.A.: 2009,
  \mnras 400, 1241.

\bibitem[{{Chevalier}(2004)}]{Chevalier04a}
{Chevalier}, R.A.: 2004, Advances in Space Research 33, 456.

\bibitem[{{Chevalier} and {Fransson}(1992)}]{Chevalier-Fransson92a}
{Chevalier}, R.A. and {Fransson}, C.: 1992, \apj 395, 540.

\bibitem[{{Chiueh} et~al.(1998) {Chiueh},{Li}, and
    {Begelman}}]{Chiueh_Li+98a} 
{Chiueh}, T., {Li}, Z.-Y., and {Begelman}, M.~C., 1998, \apj, 505, 835 

\bibitem[{{Contopoulos} et~al.(1999){Contopoulos}, {Kazanas}, and
  {Fendt}}]{Contopoulos-Kazanas+99a}
{Contopoulos}, I., {Kazanas}, D., and {Fendt}, C.: 1999, \apj 511, 351.

\bibitem[{{Cordes} et~al.(1993){Cordes}, {Romani}, and
  {Lundgren}}]{Cordes-Romani+93a}
{Cordes}, J.M., {Romani}, R.W., and {Lundgren}, S.C.: 1993, \nat 362, 133.

\bibitem[{{Dall'Osso} et~al.(2011){Dall'Osso}, {Stratta}, {Guetta}, {Covino},
  {De Cesare}, and {Stella}}]{Dall'Osso-Stratta+11a}
{Dall'Osso}, S., {Stratta}, G., {Guetta}, D., {Covino}, S., {De Cesare}, G.,
  and {Stella}, L.: 2011, \aap 526, A121.

\bibitem[{{de Jager} et~al.(2008){de Jager}, {Ferreira}, and
  {Djannati-Ata{\"i}}}]{de_Jager-Fereira+08a}
{de Jager}, O.C., {Ferreira}, S.E.S., and {Djannati-Ata{\"i}}, A.: 2008, in
  F.A. {Aharonian}, W.~{Hofmann}, and F.~{Rieger}, eds., American Institute of
  Physics Conference Series, vol. 1085 of \emph{American Institute of Physics
  Conference Series}, 199--202.

\bibitem[{{Del Zanna} et~al.(2004){Del Zanna}, {Amato}, and
  {Bucciantini}}]{Del_Zanna-Amato+04a}
{Del Zanna}, L., {Amato}, E., and {Bucciantini}, N.: 2004, \aap 421, 1063.

\bibitem[{{Del Zanna} et~al.(2003){Del Zanna}, {Bucciantini}, and
  {Londrillo}}]{Del_Zanna-Bucciantini+03a}
{Del Zanna}, L., {Bucciantini}, N., and {Londrillo}, P.: 2003, \aap 400, 397.

\bibitem[{{Del Zanna} et~al.(2006){Del Zanna}, {Volpi}, {Amato}, and
  {Bucciantini}}]{Del_Zanna-Volpi+06a}
{Del Zanna}, L., {Volpi}, D., {Amato}, E., and {Bucciantini}, N.: 2006, \aap
  453, 621.

\bibitem[{{Emmering} and {Chevalier}(1987)}]{Emmering_Chevalier87a}
{Emmering}, R.T. and {Chevalier}, R.A.: 1987, \apj 321, 334.

\bibitem[{{Ferreira} and {de Jager}(2008)}]{Ferreira-De_Jager08a}
{Ferreira}, S.E.S. and {de Jager}, O.C.: 2008, \aap 478, 17.

\bibitem[{{Gaensler} et~al.(2002){Gaensler}, {Arons}, {Kaspi}, {Pivovaroff},
  {Kawai}, and {Tamura}}]{Gaensler-Arons+02a}
{Gaensler}, B.M., {Arons}, J., {Kaspi}, V.M., {Pivovaroff}, M.J., {Kawai}, N.,
  and {Tamura}, K.: 2002, \apj 569, 878.

\bibitem[{{Gaensler} and {Slane}(2006)}]{Gaensler_Slane06a}
{Gaensler}, B.M. and {Slane}, P.O.: 2006, \araa 44, 17.

\bibitem[{{Gaensler} et~al.(2004){Gaensler}, {van der Swaluw}, {Camilo},
  {Kaspi}, {Baganoff}, {Yusef-Zadeh}, and
  {Manchester}}]{Gaensler-van_der_Swaluw+04a}
{Gaensler}, B.M., {van der Swaluw}, E., {Camilo}, F., {Kaspi}, V.M.,
  {Baganoff}, F.K., {Yusef-Zadeh}, F., and {Manchester}, R.N.: 2004, \apj 616,
  383.

\bibitem[{{Gelfand} et~al.(2009){Gelfand}, {Slane}, and
  {Zhang}}]{Gelfand_Slane+09a}
{Gelfand}, J.D., {Slane}, P.O., and {Zhang}, W.: 2009, \apj 703, 2051.

\bibitem[{{Giuliani}(1982)}]{Giuliani82a}
{Giuliani}, Jr., J.L.: 1982, \apj 256, 624.

\bibitem[{{Goldreich} and {Julian}(1969)}]{Goldreich_Julian69a}
{Goldreich}, P. and {Julian}, W.H.: 1969, \apj 157, 869.

\bibitem[{{Goldreich} and {Julian}(1970)}]{Goldreich-Julian70a}
{Goldreich}, P. and {Julian}, W.H.: 1970, \apj 160, 971.

\bibitem[{{Harding}(1996)}]{Harding96a}
{Harding}, A.K.: 1996, \ssr 75, 257.

\bibitem[{{Hester}(2008)}]{Hester08a}
{Hester}, J.J.: 2008, \araa 46, 127.

\bibitem[{{Hester} et~al.(1995){Hester}, {Scowen}, {Sankrit}
  et~al.}]{Hester-Scowen+95a}
{Hester}, J.J., {Scowen}, P.A., {Sankrit}, R., et~al.: 1995, \apj 448, 240.

\bibitem[{{Hibschman} and {Arons}(2001)}]{Hibschman_Arons01a}
{Hibschman}, J.A. and {Arons}, J.: 2001, \apj 560, 871.

\bibitem[{{Jun}(1998)}]{Jun98a}
{Jun}, B.I.: 1998, \apj 499, 282.

\bibitem[{{Kargaltsev} et~al.(2008){Kargaltsev}, {Misanovic}, {Pavlov}, {Wong},
  and {Garmire}}]{Kargaltsev-Misanovic+08a}
{Kargaltsev}, O., {Misanovic}, Z., {Pavlov}, G.G., {Wong}, J.A., and {Garmire},
  G.P.: 2008, \apj 684, 542.

\bibitem[{{Kargaltsev} and {Pavlov}(2010)}]{Kargaltsev_Pavlov10a}
{Kargaltsev}, O. and {Pavlov}, G.G.: 2010, X-ray Astronomy 2009; Present
  Status, Multi-Wavelength Approach and Future Perspectives 1248, 25.

\bibitem[{{Kennel} and {Coroniti}(1984{\natexlab{a}})}]{Kennel_Coroniti84a}
{Kennel}, C.F. and {Coroniti}, F.V.: 1984{\natexlab{a}}, \apj 283, 694.

\bibitem[{{Kennel} and {Coroniti}(1984{\natexlab{b}})}]{Kennel_Coroniti84b}
{Kennel}, C.F. and {Coroniti}, F.V.: 1984{\natexlab{b}}, \apj 283, 710.

\bibitem[{{Komissarov}(1999)}]{Komissarov99a}
{Komissarov}, S.S.: 1999, \mnras 303, 343.

\bibitem[{{Komissarov}(2006)}]{Komissarov06a}
{Komissarov}, S.S.: 2006, \mnras 367, 19.

\bibitem[{{Komissarov} and {Lyubarsky}(2004)}]{Komissarov-Lyubarsky04a}
{Komissarov}, S.S. and {Lyubarsky}, Y.E.: 2004, \mnras 349, 779.

\bibitem[{{Lu} et~al.(2002){Lu}, {Wang}, {Aschenbach}, {Durouchoux}, and
  {Song}}]{Lu-Wang+02a}
{Lu}, F.J., {Wang}, Q.D., {Aschenbach}, B., {Durouchoux}, P., and {Song}, L.M.:
  2002, \apjl 568, L49.

\bibitem[{{Lyubarsky} and {Liverts}(2008)}]{Lyubarsky_Livers08a}
{Lyubarsky}, Y. and {Liverts}, M.: 2008, \apj 682, 1436.

\bibitem[{{Lyubarsky}(2002)}]{Lyubarsky02a}
{Lyubarsky}, Y.E.: 2002, \mnras 329, L34.

\bibitem[{{Lyubarsky}(2003)}]{Lyubarsky03a}
{Lyubarsky}, Y.E.: 2003, \mnras 345, 153.

\bibitem[{{Mart{\'{\i}}n} et~al.(2012){Mart{\'{\i}}n}, {Torres}, and
  {Rea}}]{Matin_Torres+12a}
{Mart{\'{\i}}n}, J., {Torres}, D.F., and {Rea}, N.: 2012, \mnras 427, 415.

\bibitem[{{Melatos}(1998)}]{Melatos98a}
{Melatos}, A.\ 1998, MemSAIt, 69, 1009 


\bibitem[{{Michel}(1969)}]{Michel69a}
{Michel}, F.C.: 1969, \apj 158, 727.

\bibitem[{{Mizuno} et~al.(2011){Mizuno}, {Lyubarsky}, {Nishikawa}, and
  {Hardee}}]{Mizuno-Lyubarsky+11a}
{Mizuno}, Y., {Lyubarsky}, Y., {Nishikawa}, K.I., and {Hardee}, P.E.: 2011,
  \apj 728, 90.

\bibitem[{{Mizuno} et~al.(2012){Mizuno}, {Lyubarsky}, {Nishikawa}, and
  {Hardee}}]{Mizuno-Lyubarsky+12a}
{Mizuno}, Y., {Lyubarsky}, Y., {Nishikawa}, K.I., and {Hardee}, P.E.: 2012,
  \apj 757, 16.

\bibitem[{{Mori} et~al.(2004){Mori}, {Burrows}, {Pavlov}, {Hester}, {Shibata},
  and {Tsunemi}}]{Mori-Burrows+04a}
{Mori}, K., {Burrows}, D.N., {Pavlov}, G.G., {Hester}, J.J., {Shibata}, S., and
  {Tsunemi}, H.: 2004, in F.~{Camilo} and B.M. {Gaensler}, eds., Young Neutron
  Stars and Their Environments, vol. 218 of \emph{IAU Symposium}, 181.

\bibitem[{{Nalewajko} and {Begelman}(2012)}]{Nalewajko-Begelman12a}
{Nalewajko}, K. and {Begelman}, M.C.: 2012, \mnras 427, 2480.

\bibitem[{{Ng} et~al.(2012){Ng}, {Bucciantini}, {Gaensler}, {Camilo},
  {Chatterjee}, and {Bouchard}}]{Ng-Bucciantini+12a}
{Ng}, C.Y., {Bucciantini}, N., {Gaensler}, B.M., {Camilo}, F., {Chatterjee},
  S., and {Bouchard}, A.: 2012, \apj 746, 105.

\bibitem[{{O'Neill} et~al.(2012){O'Neill}, {Beckwith}, and
  {Begelman}}]{O'Neill-Beckwith+12a}
{O'Neill}, S.M., {Beckwith}, K., and {Begelman}, M.C.: 2012, \mnras 422, 1436.

\bibitem[{{Pacini} and {Salvati}(1973)}]{Pacini_Salvati73a}
{Pacini}, F. and {Salvati}, M.: 1973, \apj 186, 249.

\bibitem[{{Pavlov} et~al.(2006){Pavlov}, {Sanwal}, and
  {Zavlin}}]{Pavlov-Sanwal+06a}
{Pavlov}, G.G., {Sanwal}, D., and {Zavlin}, V.E.: 2006, \apj 643, 1146.

\bibitem[{{Pavlov} et~al.(2003){Pavlov}, {Teter}, {Kargaltsev}, and
  {Sanwal}}]{Pavlov-Teter+03a}
{Pavlov}, G.G., {Teter}, M.A., {Kargaltsev}, O., and {Sanwal}, D.: 2003, \apj
  591, 1157.

\bibitem[{{Porth} et~al.(2013){Porth}, {Komissarov}, and
  {Keppens}}]{Porth-Komissarov+13a}
{Porth}, O., {Komissarov}, S.S., and {Keppens}, R.: 2013, \mnras 431, L48.

\bibitem[{{Rees} and {Gunn}(1974)}]{Rees_Gunn73a}
{Rees}, M.J. and {Gunn}, J.E.: 1974, \mnras 167, 1.

\bibitem[{{Reynolds}(2004)}]{Reynolds04a}
{Reynolds}, S.: 2004, in J.P. {Paill{\'e}}, ed., 35th COSPAR Scientific
  Assembly, vol.~35 of \emph{COSPAR Meeting}, 3572.

\bibitem[{{Reynolds} and {Chanan}(1984)}]{Reynolds_Chanan84a}
{Reynolds}, S.P. and {Chanan}, G.A.: 1984, \apj 281, 673.

\bibitem[{{Reynolds} and {Chevalier}(1984)}]{Reynolds_Chevalier84a}
{Reynolds}, S.P. and {Chevalier}, R.A.: 1984, \apj 278, 630.

\bibitem[{{Romani} and {Ng}(2003)}]{Romani-Ng03a}
{Romani}, R.W. and {Ng}, C.Y.: 2003, \apjl 585, L41.

\bibitem[{{Romani} et~al.(2005){Romani}, {Ng}, {Dodson}, and
  {Brisken}}]{Romani-Ng+05a}
{Romani}, R.W., {Ng}, C.Y., {Dodson}, R., and {Brisken}, W.: 2005, \apj 631,
  480.

\bibitem[{{Schweizer} et~al.(2013){Schweizer}, {Bucciantini}, {Idec},
  {Nilsson}, {Tennant}, {Weisskopf}, and {Zanin}}]{Schweizer-Bucciantini+13a}
{Schweizer}, T., {Bucciantini}, N., {Idec}, W., {Nilsson}, K., {Tennant}, A.,
  {Weisskopf}, M., and {Zanin}, R.: 2013, ArXiv e-prints .

\bibitem[{{Sironi} and {Spitkovsky}(2009)}]{Sironi_Spitkovsky09a}
{Sironi}, L. and {Spitkovsky}, A.: 2009, \apj 698, 1523.

\bibitem[{{Sironi} and {Spitkovsky}(2011{\natexlab{a}})}]{Sironi_Spitkovsky11a}
{Sironi}, L. and {Spitkovsky}, A.: 2011{\natexlab{a}}, \apj 741, 39.

\bibitem[{{Sironi} and {Spitkovsky}(2011{\natexlab{b}})}]{Sironi_Spitkovsky11b}
{Sironi}, L. and {Spitkovsky}, A.: 2011{\natexlab{b}}, \apj 726, 75.

\bibitem[{{Slane} et~al.(2004){Slane}, {Helfand}, {van der Swaluw}, and
  {Murray}}]{Slane-Helfand+04a}
{Slane}, P., {Helfand}, D.J., {van der Swaluw}, E., and {Murray}, S.S.: 2004,
  \apj 616, 403.

\bibitem[{{Smith} et~al.(1988){Smith}, {Jones}, {Dick}, and
  {Pike}}]{Smith-Jones+88a}
{Smith}, F.G., {Jones}, D.H.P., {Dick}, J.S.B., and {Pike}, C.D.: 1988, \mnras
  233, 305.

\bibitem[{{Spitkovsky}(2006)}]{Spitkovsky06a}
{Spitkovsky}, A.: 2006, \apjl 648, L51.

\bibitem[{{Tanaka} and {Takahara}(2010)}]{Tanaka_Takahara10a}
{Tanaka}, S.J. and {Takahara}, F.: 2010, \apj 715, 1248.

\bibitem[{{Tanaka} and {Takahara}(2013)}]{Tanaka_Takahara13a}
{Tanaka}, S.J. and {Takahara}, F.: 2013, \mnras 429, 2945.

\bibitem[{{Timokhin}(2006)}]{Timokhin06a}
{Timokhin}, A.~N.: 2006, \mnras, 368, 1055 

\bibitem[{{Torres} et~al.(2013){Torres}, {Cillis}, and {Mart{\'{\i}}n
  Rodriguez}}]{Torres_Cillis+13a}
{Torres}, D.F., {Cillis}, A.N., and {Mart{\'{\i}}n Rodriguez}, J.: 2013, \apjl
  763, L4.

\bibitem[{{van der Swaluw}(2003)}]{van_der_Swaluw03a}
{van der Swaluw}, E.: 2003, \aap 404, 939.

\bibitem[{{van der Swaluw} et~al.(2002){van der Swaluw}, {Achterberg}, and
  {Gallant}}]{van_der_Swaluw-Achterberg+02a}
{van der Swaluw}, E., {Achterberg}, A., and {Gallant}, Y.A.: 2002, in P.O.
  {Slane} and B.M. {Gaensler}, eds., Neutron Stars in Supernova Remnants, vol.
  271 of \emph{Astronomical Society of the Pacific Conference Series}, 135.

\bibitem[{{van der Swaluw} et~al.(2001){van der Swaluw}, {Achterberg},
  {Gallant}, and {T{\'o}th}}]{van_der_Swaluw-Achterberg+01a}
{van der Swaluw}, E., {Achterberg}, A., {Gallant}, Y.A., and {T{\'o}th}, G.:
  2001, \aap 380, 309.

\bibitem[{{van der Swaluw} et~al.(2004){van der Swaluw}, {Downes}, and
  {Keegan}}]{van_der_Swaluw-Downes+04a}
{van der Swaluw}, E., {Downes}, T.P., and {Keegan}, R.: 2004, \aap 420, 937.

\bibitem[{{Velusamy}(1985)}]{Velusamy85a}
{Velusamy}, T.: 1985, \mnras 212, 359.

\bibitem[{{Vigelius} et~al.(2007){Vigelius}, {Melatos}, {Chatterjee},
  {Gaensler}, and {Ghavamian}}]{Vigelius-Melatos+07a}
{Vigelius}, M., {Melatos}, A., {Chatterjee}, S., {Gaensler}, B.M., and
  {Ghavamian}, P.: 2007, \mnras 374, 793.

\bibitem[{{Volpi} et~al.(2008){Volpi}, {Del Zanna}, {Amato}, and
  {Bucciantini}}]{Volpi-Del_Zanna+08a}
{Volpi}, D., {Del Zanna}, L., {Amato}, E., and {Bucciantini}, N.: 2008, \aap
  485, 337.

\bibitem[{{Vorster} et~al.(2013){Vorster}, {Ferreira}, {de Jager}, and
  {Djannati-Ata{\"i}}}]{Vorster-Ferreira+13a}
{Vorster}, M.J., {Ferreira}, S.E.S., {de Jager}, O.C., and {Djannati-Ata{\"i}},
  A.: 2013, \aap 551, A127.

\bibitem[{{Weiler} and {Panagia}(1978)}]{Weiler_Panagia78a}
{Weiler}, K.W. and {Panagia}, N.: 1978, \aap 70, 419.

\bibitem[{{Weisskopf} et~al.(2000){Weisskopf}, {Hester}, {Tennant}
  et~al.}]{Weisskopf-Hester+00a}
{Weisskopf}, M.C., {Hester}, J.J., {Tennant}, A.F., et~al.: 2000, \apjl 536,
  L81.

\bibitem[{{Wilkin}(1996)}]{Wilkin96a}
{Wilkin}, F.P.: 1996, \apjl 459, L31.

\bibitem[{{Wilkin}(2000)}]{Wilkin00a}
{Wilkin}, F.P.: 2000, \apj 532, 400.

\end{thebibliography}
%\bibliographystyle{prova}

\end{document}